\documentclass[prb,aps,twocolumn,showpacs]{revtex4}

\usepackage{amsmath,amsthm,amssymb}

\usepackage{amsmath,graphicx}
\usepackage{color,ulem}
\newlength{\figwidth}
\newlength{\figlarge}
\begin{document}
\title{Topological antiferromagnetic phase in a correlated Bernevig-Hughes-Zhang model}

\author{Tsuneya Yoshida}
\author{Robert Peters}
\author{Satoshi Fujimoto}
\author{Norio Kawakami}

\affiliation{
Department of Physics, Kyoto University, 
Kyoto 606-8502, Japan}

\date{\today}
\begin{abstract}
Topological properties of antiferromagnetic phases are studied for a correlated topological band insulator by applying the dynamical mean field theory to an extended Bernevig-Hughes-Zhang model including the Hubbard interaction. The calculation of the magnetic moment and the spin Chern number confirms the existence of a non-trivial antiferromagnetic (AF) phase beyond the Hartree-Fock theory. In particular, we uncover the intriguing fact that the topologically non-trivial AF phase is essentially stabilized by correlation effects but not by the Hartree shifts alone. This counterintuitive effect is demonstrated, through a comparison with the Hartree-Fock results, and should apply for generic topological insulators with strong correlations.
\end{abstract}
\maketitle

\section{Introduction}

In the last three decades, a new class of quantum phases which are described by topological field theory has attracted much interest. The topological structure is reflected in the existence of gapless edge states and in characteristic magnetoelectric responses. The first example was an integer quantum Hall phase, which is characterized by the Thouless-Kohmoto-Nightingale-de Nijs (TKNN) number and shows the quantized Hall conductivity in an applied magnetic field.\cite{TKNN82} Furthermore, in this decade, $\mathbb{Z}_2$ topological band insulators (TBIs) have been found as a new family of topological phases.\cite{Hasan10,Qi10,Bernevig06,Kane05_1,Kane05_2,Bernevig06_BHZ,3DTI_Roy,3DTI_Moore,3DTI_Fu07,TBI_3DFu07,Qi08} For the realization of a $\mathbb{Z}_2$ topological phase, the spin-orbit (SO) coupling plays the essential role instead of the magnetic field. TBIs have been extensively studied from both experimental and theoretical viewpoints, and their realization in two- and three-dimensional systems has been found in $\mathrm{HgTe}/\mathrm{CdTe}$ quantum wells \cite{exp_2D-QW_MKonig} and bismuth based compounds.\cite{exp_3D-bismuth_DHsieh,exp_3D-bismuth_YXia,exp_3D-bismuth_YLChen}

Recently, correlation effects on the TBIs have spurred intense activities in theory, since electron correlations are expected to create exotic states under topologically non-trivial conditions. This issue has further been stimulated by theoretical proposals for the realization of topological phases in $d$-, $f$- electron systems, (e.g. iridium oxides, \cite{NaIrO_Nagaosa09,TMI_LBalents09} Heusler compounds, \cite{Heusler_Chadov10,Heusler_Lin10} and filled skutterudites \cite{skutterudites_Yan12}).  In the context of correlation effects on TBIs, the competition between long-range ordered phases and TBIs have been addressed.\cite{Yamaji11,AFvsTBI_DQMC_Hohenadler11,AFvsTBI_DQMC_edge_early_Zheng10,AFvsTBI_VCA_Yu11,AFvsTBI_CDMFTWu11,Rachel10,SL_Griset,Varney10,Wang10} For example, a competition between an antiferromagnetic (AF) phase and the TBI phase was studied by numerical approaches as well as mean field theory.  In a quantum Monte Carlo study by Hohenadler \textit{et al.}, \cite{AFvsTBI_DQMC_Hohenadler11} it was clarified that the TBI phase can change into a topologically trivial AF phase. This statement was supported by the results obtained with a variational cluster approach and also by the results obtained in a cluster dynamical mean field study.\cite{AFvsTBI_VCA_Yu11,AFvsTBI_CDMFTWu11}  The former study also clarified the absence of edge states in the AF phase.\cite{AFvsTBI_VCA_Yu11}

In parallel with these efforts, antiferromagnetic topological insulators (AFTIs), i.e. AF insulators having non-trivial topological structures, have been studied.\cite{AFTI_Mong10,AFTIB_Essin11,AFTBI_Guo11,AFTBI_He11,AFTBI_He12} Such a phase was first proposed by Mong \textit{et al.}, \cite{AFTI_Mong10} who found that if a three-dimensional insulator breaks both time-reversal ($\Theta$) and primitive-lattice translational ($T_{1/2}$) symmetries but preserves the combination $S= \Theta T_{1/2}$ ($S$ serves as the time-reversal operator at a certain plane of the Brillouin zone), then the system may have a topological structure. According to the periodic table of TBIs,\cite{periodic_table_Schnyder,periodic_table_Kitaev,periodic_table_Ryu} AFTIs may not be realized in generic two-dimensional systems, but we can still find such AFTI phases under spin quantized conditions.\cite{AFTBI_Guo11}

Because the above studies have been restricted to Hartree-Fock mean-field theory, there are still open and interesting questions about AFTIs.
 This paper aims to study the AFTIs by incorporating electron correlations beyond the Hartree-Fock treatment. To this end, we analyze an extended Bernevig-Hughes-Zhang (BHZ) model\cite{Bernevig06_BHZ} including the Hubbard interaction with the dynamical mean field theory (DMFT) and the numerical renormalization group (NRG). We elucidate the remarkable fact that an AF phase can have a topologically non-trivial structure near the transition point even if the Hartree-Fock treatment does not support it. We demonstrate that the renormalization of the electronic states due to strong correlations is essential to stabilize the AFTI phase. This kind of renormalization is inherent and ubiquitous in correlated TBIs, and may play the key role in realizing non-trivial ordered phases.

This paper is organized as follows. First, the model and the method are briefly explained in the next section. In Sec. \ref{sec. numerical results}, we present the results obtained with the DMFT, and discuss how the AFTI is stabilized in the presence of strong correlation effects. A brief summary is given in the last section.

\section{Model and Method}\label{sec. model and method}
We extend the BHZ model, which is defined on a two-dimensional square lattice, to include an on-site Coulomb interaction.\cite{TBI_Mott_Yoshida,TBI_Mott_Tada} The Hamiltonian reads
\begin{widetext}
\begin{eqnarray}
H &=& H_{BHZ} +U\sum_{i,\alpha} n_{i,\alpha,\uparrow} n_{i,\alpha,\downarrow}\\
  H_{BHZ}&=& \sum_{i,\alpha,\sigma} \varepsilon_\alpha n_{i,\alpha,\sigma} -\sum_{ i,j ,\sigma} \hat{c}^{\dagger}_{i,\alpha,\sigma} \hat{t}_{i,j,\sigma,\alpha,\alpha'} \hat{c}_{j,\alpha',\sigma},\\
-\hat{t}_{i,j \sigma} &=&\left(
\begin{array}{cc}
 -t_{1}(\delta_{i,j\pm \hat{x}}+\delta_{i,j\pm \hat{y}})& t'(-\mathrm{sign}(\sigma)(\delta_{i,j+\hat{y}}-\delta_{i,j-\hat{y}})+i(\delta_{i,j+\hat{x}}-\delta_{i,j-\hat{x}}))  \\
t'(\mathrm{sign}(\sigma)(\delta_{i,j+\hat{y}}-\delta_{i,j-\hat{y}})+i(\delta_{i,j+\hat{x}}-\delta_{i,j-\hat{x}})) & t_{2}(\delta_{i,j\pm \hat{x}}+\delta_{i,j\pm \hat{y}})
\end{array}
\right),
\end{eqnarray}
\end{widetext}
where $n_{i,\alpha,\sigma}=c^{\dagger}_{i,\alpha,\sigma}c_{i,\alpha,\sigma}$. The operator $c^{\dagger}_{i,\alpha,\sigma}(c_{i,\alpha,\sigma})$ creates (annihilates) an electron at site $i$ and in orbital $\alpha=1,2$ and  spin $\sigma=\uparrow, \downarrow$ state. The off-diagonal elements of the hopping matrix $\hat{t}$ drive the system into a non-trivial band insulator at $U=0$.
We analyze the system using DMFT, which treats local correlations exactly and is suitable for the systematic calculation of arbitrary strength of the Coulomb interaction. In DMFT, the original lattice problem is mapped onto an effective impurity model, which is solved self-consistently.\cite{DMFT_Georges,DMFT_Metzner,DMFT_Muller-Hartmann} 
The self-consistency equation for a paramagnetic phase is given by
\begin{eqnarray}
 \hat{g}_{\sigma}^{-1}(\omega)&=& [\sum_{{\bf k}}\frac{1}{(\omega+i\delta) \mathbb{I}-\hat{h}_{\sigma}({\bf k})-\hat{\Sigma}^{R}_{\sigma}(\omega)} ]^{-1} +\hat{\Sigma}^{R}_{\sigma}(\omega), \nonumber 
\end{eqnarray}
where $\hat{h}_{\sigma}({\bf k})$ is the Fourier transform of the hopping matrix.
The self-energy of the lattice Green's function $\hat{\Sigma}_{\sigma}(\omega)$ can be computed from the Green's function $\hat{g}(\omega)$ of the effective impurity model. 
Note that, since the orbitals are hybridized by SO coupling, the hybridization function is diagonal in the orbital index. Thus, off-diagonal elements of the self-energy $(\hat{\Sigma}_{\sigma})_{\alpha,\bar{\alpha}}$ are supposed to vanish in our model.
In this study, we employ the NRG method to solve the impurity model,\cite{NRG_Wilson,NRG_Bulla} which is a powerful method for calculations at zero-temperature. To analyze the AF phase, we divide the original square lattice into two sublattices specified by checkerboard pattern.

For simplicity, we study the particle-hole symmetric case and choose the model parameters as $t_{1}=t_{2}=t$, $t'=0.1t$ and $\varepsilon_1 (\varepsilon_2)=-t (t)$. The hopping integral $t$ is chosen as the energy unit.\\

\section{Numerical Results}\label{sec. numerical results}

\begin{figure}[!h]
\begin{minipage}{0.48\hsize}
\begin{center}
\includegraphics[width=40mm, height=39mm, clip]{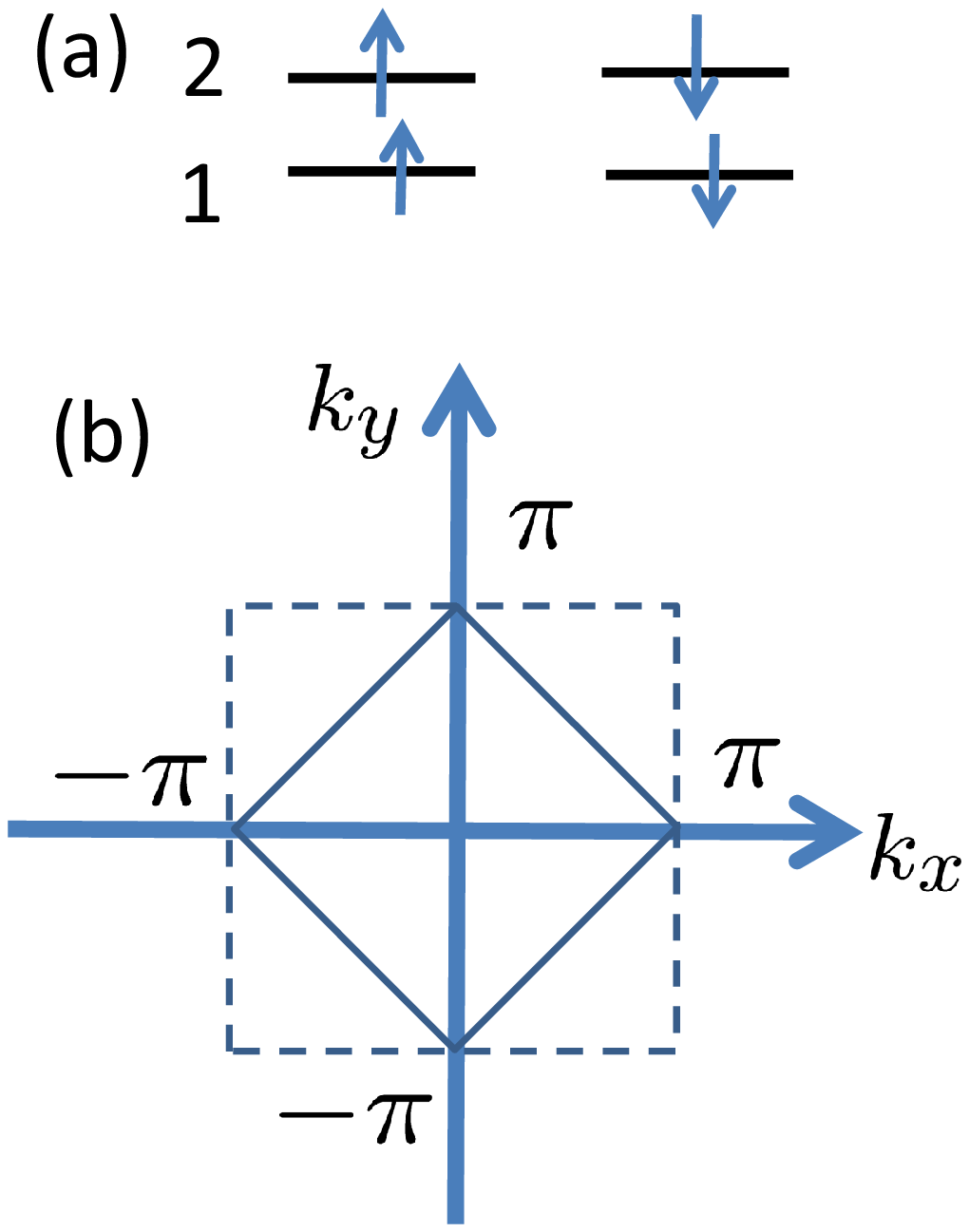}
\end{center}
\end{minipage}
\begin{minipage}{0.48\hsize}
\begin{center}
\includegraphics[width=47mm,clip]{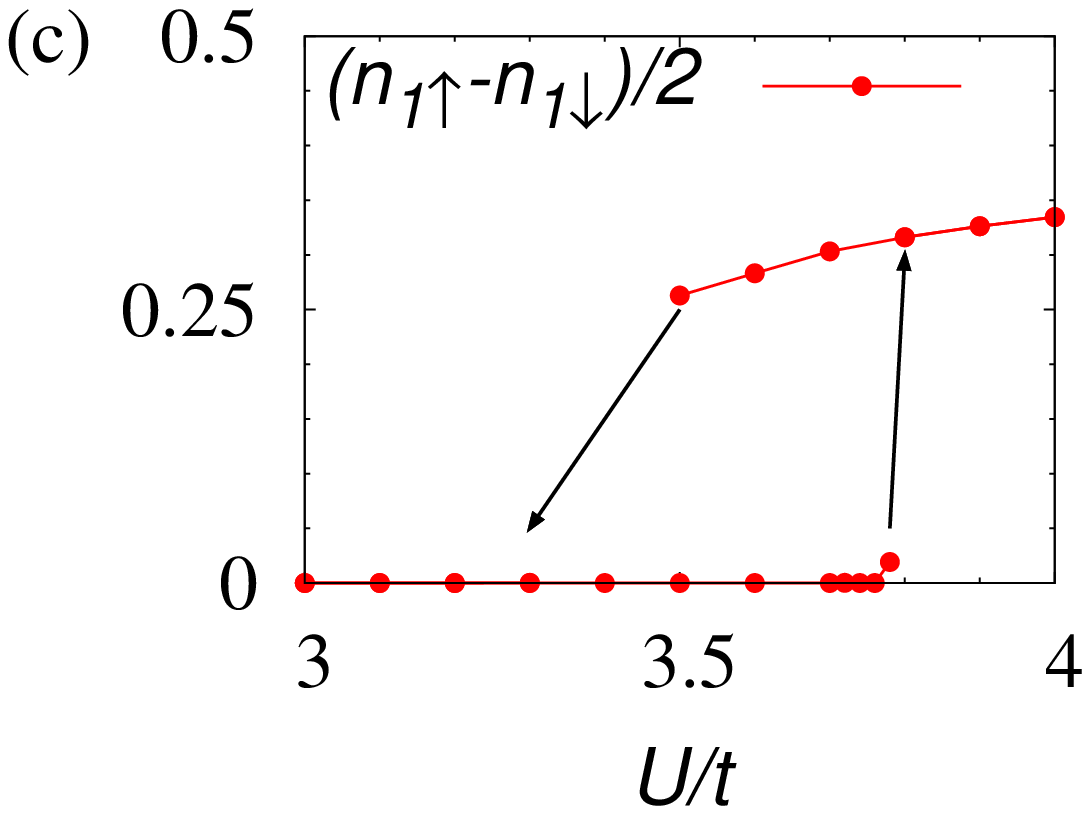}
\end{center}
\end{minipage}
\hspace{-1cm}
\begin{minipage}{0.48\hsize}
\begin{center}
\includegraphics[width=45mm,clip]{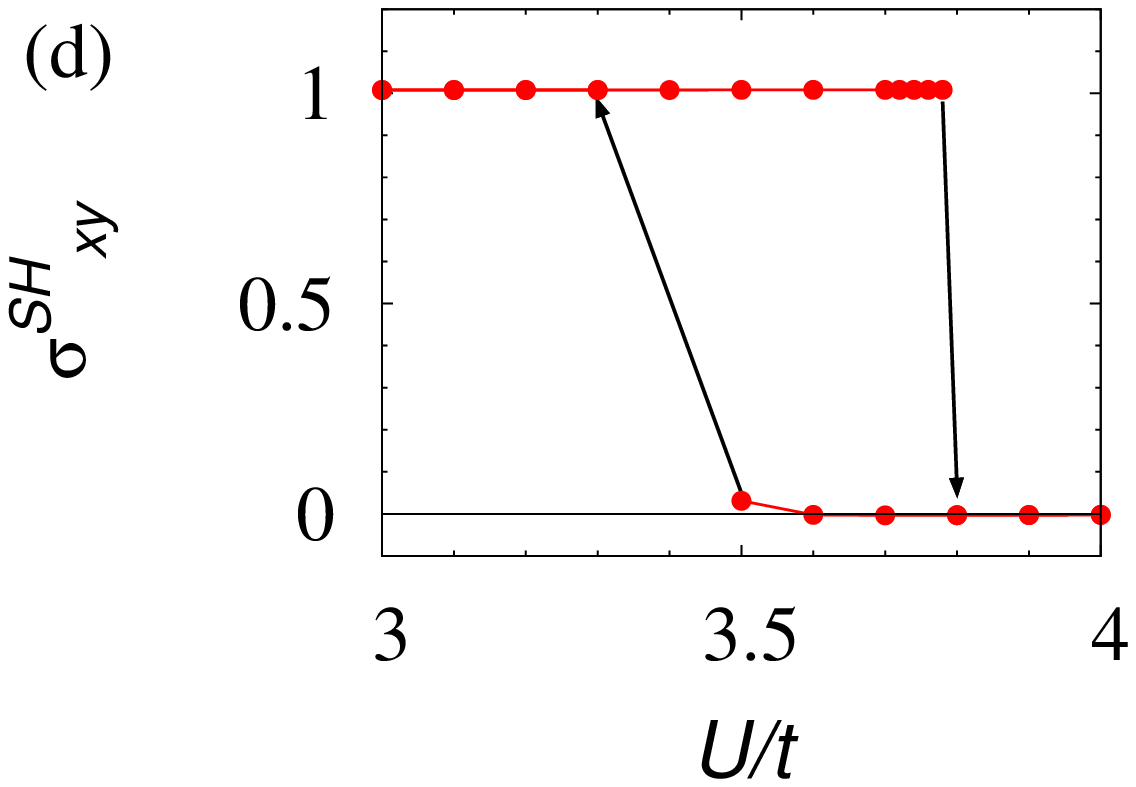}
\end{center}
\end{minipage}
\begin{minipage}{0.48\hsize}
\begin{center}
\includegraphics[width=47mm,clip]{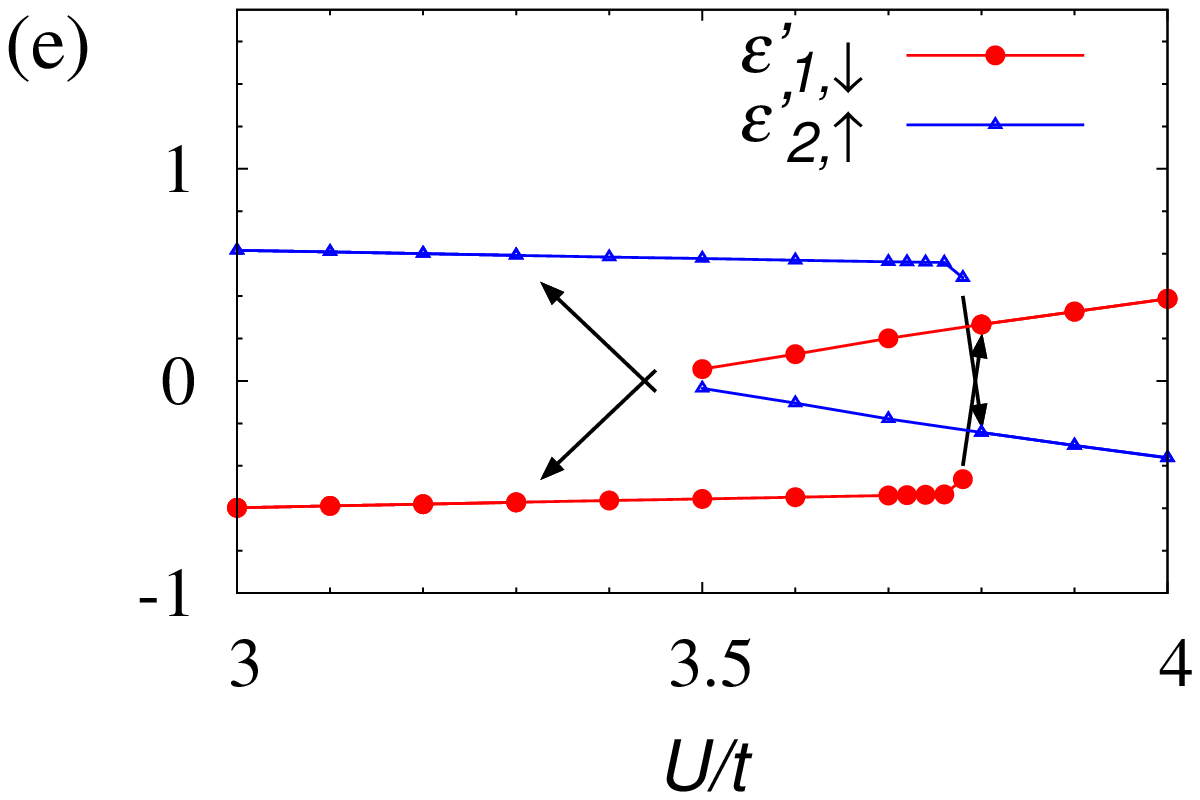}
\end{center}
\end{minipage}
\caption{
The results obtained with the Hartree-Fock approximation:
(a) spin configuration of the AF phase, (b) sketch of the reduced Brillouin zone, (c) magnetic moment at orbital 1, (d) spin-Hall conductivity, (e) energy level of each orbital modified by the Hartree shift.
For $3.5<U<3.78$, the coexistence region inherent in the first-order transition, can be found. The energy of each phase crosses at $U\sim 3.6$, which determines the phase transition point for the thermodynamically stable phase diagram.
Note that in this figure, no non-trivial AF phase is found.
}
\label{fig:Hartree}
\end{figure}

First, let us briefly discuss the Hartree-Fock results. In Fig. \ref{fig:Hartree}, the results are shown as a function of the interaction strength $U$. A close examination of the effective Hamiltonian in the strong coupling region elucidates that the off-diagonal elements of the hopping matrix $\hat{t}_{\sigma}$ induce a spin dependent exchange;
the $z$-component of the exchange interaction is antiferromagnetic, while the in-plane components are ferromagnetic (antiferromagnetic) between neighbors in $y$- ($x$-) direction, respectively. This results from the fact that the phase of the intra-orbital hopping affects the spin-exchange process.
Thus, the spin configuration of the AF phase is expected to be the one shown in Fig. \ref{fig:Hartree}(a). In this figure, the staggered moment is parallel to the $z$-axis, and the in-plane components are zero.
Here it should be noted that this effective Hamiltonian preserves four-fold symmetry; the Hamiltonian is invariant under the $\pi/2$ spatial rotation combined with the rotation of the spin space ($S^{x(y)}_{i,1} \rightarrow -S^{x(y)}_{i,1}$) at every site, where $S^{x(y)}_{i,1}$ represents the $x$ $(y)$ component of spin operator at site $i$ and orbital 1, respectively.

In Fig. \ref{fig:Hartree}(c), we show that for $U<3.3$ ($U>3.8$), the system is in a paramagnetic phase (AF phase) respectively. 
Accordingly, the Brillouin zone is reduced (Fig. \ref{fig:Hartree} (b)), since the ordering vector is ($\pi,\pi$). In this figure, for $3.5<U<3.78$, a hysteresis behavior is observed\cite{Footnote_hysteresis_HF}, and in the region of $3.3<U<3.5$, the paramagnetic (AF) solution is stable (unstable). 
Corresponding to the magnetic transition, in Fig. \ref{fig:Hartree} (d), we observe that the topological property changes at $U=3.8$ ($U=3.5$) with increasing (decreasing) interaction. As a result, a non-trivial AF phase is not found. This is attributed to the large Hartree shift induced in the AF phase. Recall that in our model, the origin of the gap depends on the energy splitting ($\epsilon_2'-\epsilon_1'$), where ${\varepsilon '}_{\alpha,\sigma}$ is the energy level of each orbital including the Hartree shift, and if it becomes zero, the gap closing occurs. In the region of $\epsilon_1'-\epsilon_2' < 0$, the gap is induced by the SO interaction. Moreover, in this region, we can confirm that the system possesses non-trivial topology (see Fig. 1 (d)). If the Green's function has no anomaly (gap closing or zeros of it), the topological properties are never changed.\cite{Volovik03,Gurarie11} 
We thus conclude that for $\epsilon_1'<\epsilon_2'$ the system is driven into the non-trivial phase. Keeping this in mind, we plot the energy levels in Fig. \ref{fig:Hartree}(e). In this figure, a region satisfying the non-trivial condition is not found within the AF phase. We have also checked other choices of  parameters, but could only find a topological-trivial AF phase in the physically sensible parameter regions.

\begin{figure}[!h]
\begin{minipage}{0.7\hsize}
\begin{center}
\includegraphics[width=\hsize]{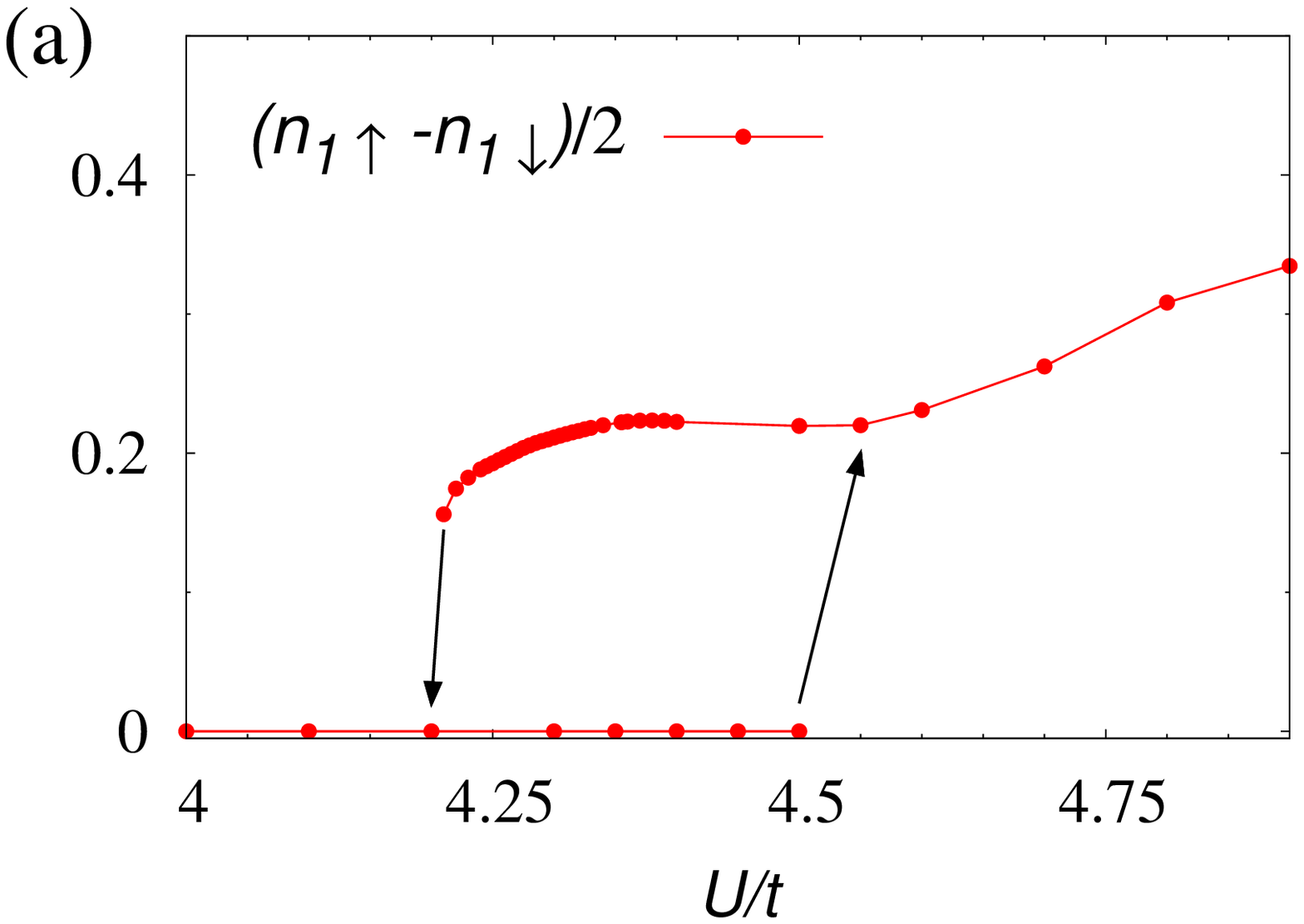}
\end{center}
\end{minipage}
\vspace{1cm}
\begin{minipage}{0.7\hsize}
\begin{center}
\includegraphics[width=\hsize]{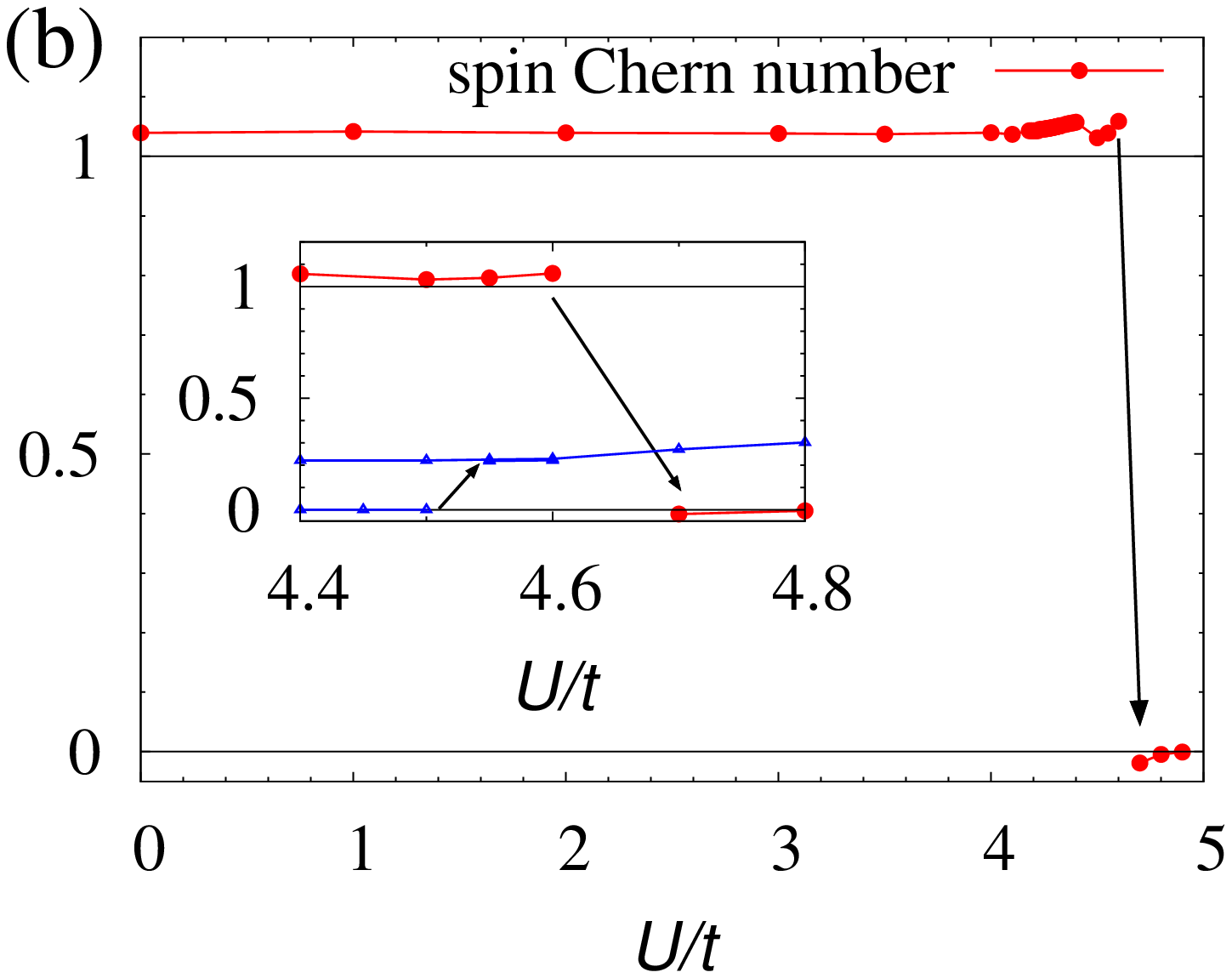}
\end{center}
\end{minipage}
\caption{(Color Online)
(a) The AF moment of orbital 1 as a function of interaction strength.
(b) The SChN as a function of interaction strength. Inset: the SChN (red) and the magnetic moment of orbital 1 (blue) are plotted.
In figure (a), the coexistence region is also observed in the DMFT results for $4.2<U<4.5$. Here the first-order phase transition point, where the energy of each phase crosses, is estimated as $U\sim 4.3$.
}
\label{fig:M}
\end{figure}

Let us now discuss the results obtained with DMFT, and clarify what happens when electron correlations are taken into account. In Fig. \ref{fig:M}(a), the AF moment is plotted as a function of interaction strength. We can clearly observe a jump at $U=4.5$ ($U=4.2$) with increasing (decreasing) interaction strength, respectively. 
In order to clarify the topological properties, we calculate the spin Chern number (SChN).\cite{Volovik03,Sheng05,Fu06,Fukui07} 
We note again that the AFTI phase can be stabilized in our model, since the Green's function for the magnetic phase is diagonal in spin space, and therefore its topological properties are specified by the SChN even in this phase.\cite{AFTBI_Guo11}
\begin{eqnarray}
N_{\mathrm{SChN}} &=& \frac{\epsilon^{\mu \nu \rho}}{48\pi^2} \int d^3p \sum_{\sigma} \mathrm{sign}(\sigma) \nonumber \\ 
 &&  \mathrm{tr}[ \hat{G}^{-1}_{\sigma}(p) \frac{\partial \hat{G}_{\sigma}(p)}{\partial p_{\mu}}  \hat{G}^{-1}_{\sigma}(p) \frac{\partial \hat{G}_{\sigma}(p)}{\partial p_{\nu}} \hat{G}^{-1}_{\sigma}(p) \frac{\partial \hat{G}_{\sigma}(p)}{\partial p_{\rho}}   ]. \nonumber
\end{eqnarray}
Here, the notation $p=(i\omega,\mathbf{p})$ is used. Note that even in the interacting case, this quantity is proportional to the spin-Hall conductivity. \cite{Ishikawa87,Haldnane04,Footnote_QSH}
In Fig. \ref{fig:M}(b), the SChN is plotted as a function of interaction strength. Note that the system has a bulk gap for each phase, as discussed momentarily below.
In this figure, we find that the system is in the TBI phase at $U=0$ and retains its topological properties up to $U=4.6$. Moreover, as seen in Fig. \ref{fig:M} (a), for $4.5<U$, only the AF solution is obtained. We therefore end up with an AFTI phase, which is not observed in the Hartree-Fock treatment. Furthermore, in the coexistence region, the AFTI phase is also stabilized for $4.3 \lesssim U <4.5$. We thus come to the remarkable conclusion that strong correlations (quantum fluctuations) drive the system to the AFTI phase. Here, it should be noted that, in the AFTI phase, we can observe both the quantized spin-Hall conductivity and the AF order.

\begin{figure}[!h]
\begin{center}
\includegraphics[width=80mm,clip]{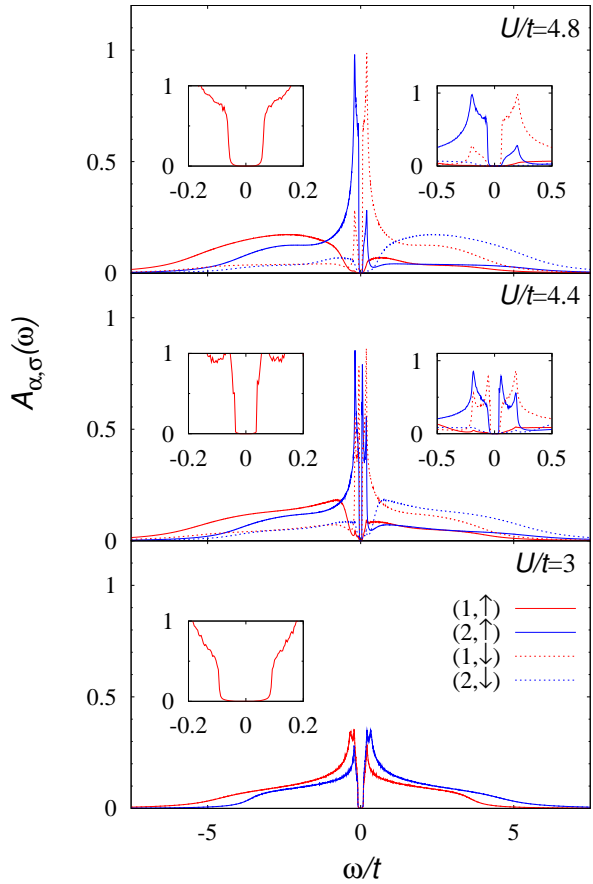}
\end{center}
\caption{(Color Online)
Spectral functions for an up-spin dominant sublattice ($A_{\alpha,\sigma}(\omega)$) for several values of interaction strength. At $U=3$, $(4.4,4.8)$, the system is in the TBI, (non-trivial AF, trivial AF) phase respectively.
Solid red (solid blue, dashed red, dashed blue) line represents that of the state $(\alpha,\sigma)=(1,\uparrow)$ ($(1, \downarrow), (2,\uparrow), (2,\downarrow)$) respectively. 
Inset (left): the total spectral functions ($\Sigma_{\alpha,\sigma} A_{\alpha,\sigma}$) near the Fermi energy ($\omega=0$). Inset (right): the spectral function near the Fermi energy  around $\omega=0$.
}
\label{fig:total_DOS}
\end{figure}

To understand how the electronic states are changed at the transition points, we now discuss spectral properties. The local density of states, defined as $A_{\alpha,\sigma}(\omega)   =-\frac{1}{\pi} \mathrm{Im} \hat{G}_{\alpha,\alpha,\sigma} ( \omega + i \delta )  $, for an up-spin dominant sublattice is shown in Fig. \ref{fig:total_DOS} for several values of interaction strength.
In the insets, the total spectral function ($A(\omega)=\sum_{\alpha,\sigma} A_{\alpha,\sigma}(\omega)$) near the Fermi energy is plotted.  We observe a bulk gap in both phases.
In the paramagnetic phase, the spectral functions of each spin state are identical.
In the AFTI phase ($U=4.4$), the spectral functions for ($\alpha,\sigma$)=($1,\downarrow$) and ($2,\uparrow$) have peaks near the Fermi energy, while those for ($\alpha,\sigma$)=($1,\uparrow$) and ($2,\downarrow$) have a hump structure around $\omega \sim \pm U/2$.
With further increasing the interaction strength, the topological structure of this system becomes trivial. In this trivial AF phase, we can see that the electronic structure near the Fermi energy is changed; the peak just below (above) the Fermi energy is mainly composed of the state $(\alpha,\sigma)=(2,\uparrow)$ ( $(1,\downarrow)$ )  respectively.

\begin{figure}[!h]
\begin{minipage}{1\hsize}
\begin{center}
\includegraphics[width=80mm,clip]{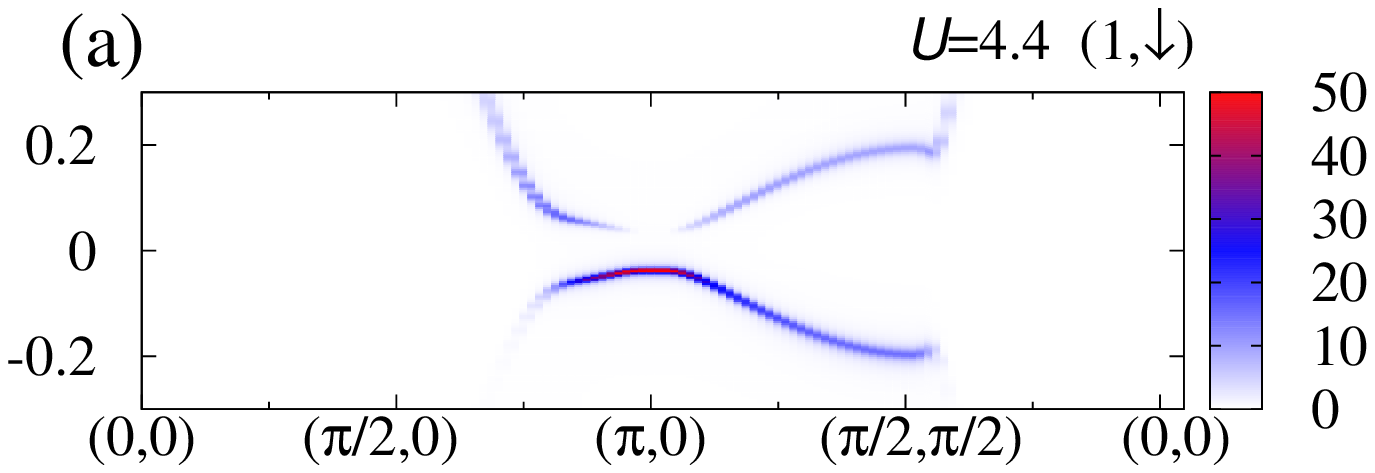}
\end{center}
\end{minipage}
\begin{minipage}{1\hsize}
\begin{center}
\includegraphics[width=80mm,clip]{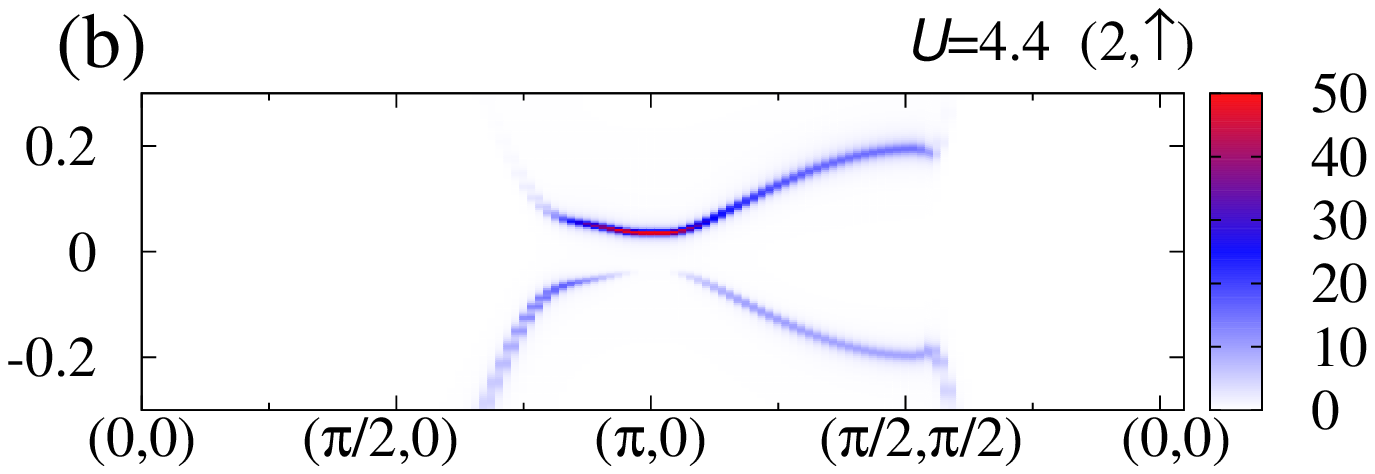}
\end{center}
\end{minipage}
\begin{minipage}{1\hsize}
\begin{center}
\includegraphics[width=80mm,clip]{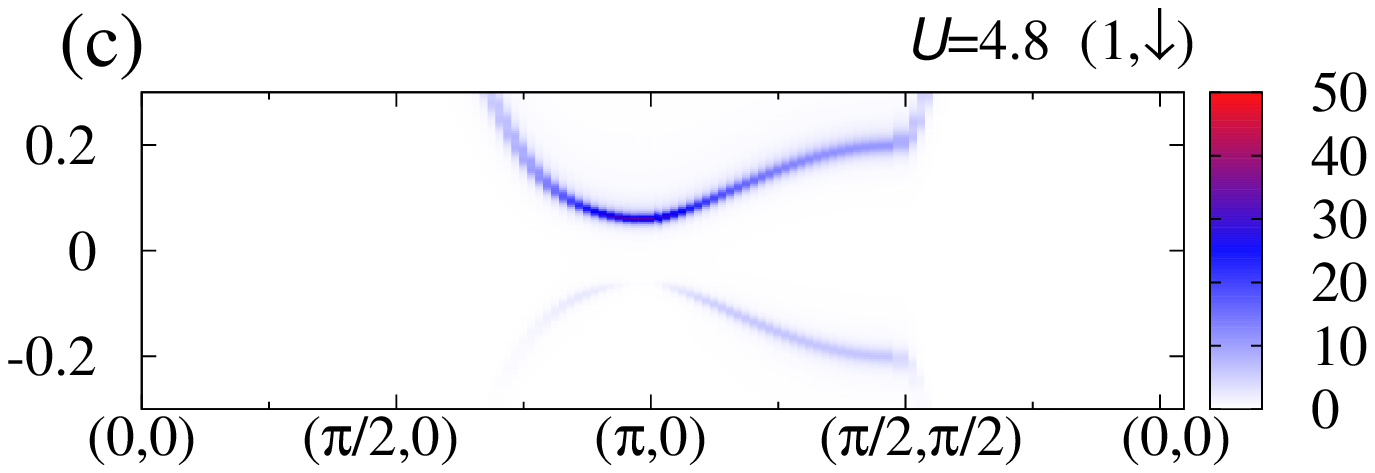}
\end{center}
\end{minipage}
\begin{minipage}{1\hsize}
\begin{center}
\includegraphics[width=80mm,clip]{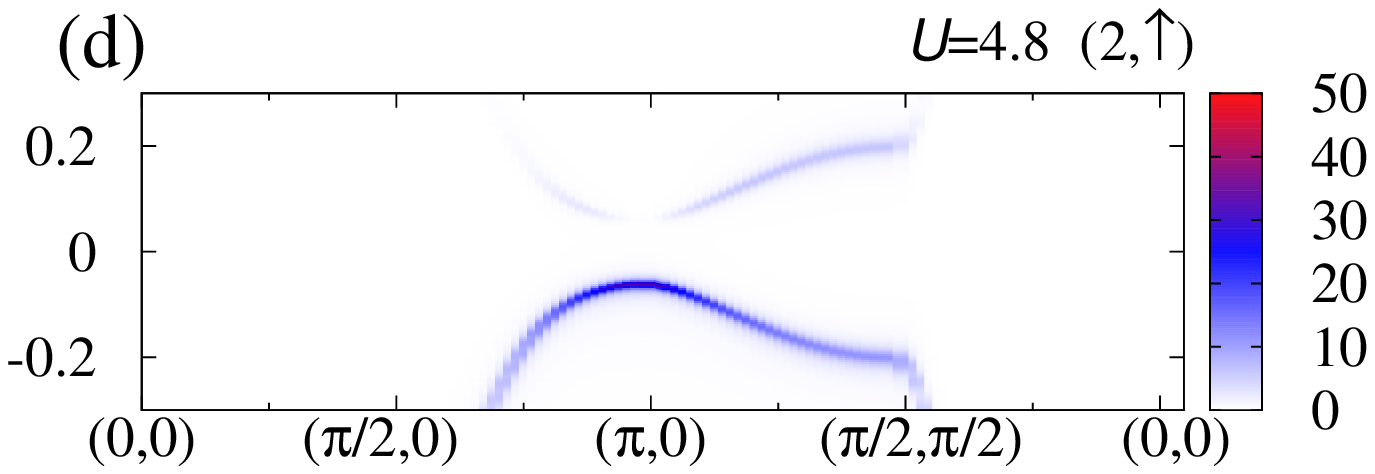}
\end{center}
\end{minipage}
\caption{
Momentum resolved spectral functions ($A_{\alpha, \sigma}(\mathbf{k},\omega)$) for $(\alpha,\sigma)=(1,\downarrow),(2,\uparrow)$:
(a),(b) for $U=4.4$, (c),(d) for $U=4.8$.
}
\label{fig:Ak}
\end{figure}

For further insights into the AFTI phase, the momentum-resolved spectral functions are plotted in Fig. \ref{fig:Ak}. In these figures, it is shown that states near the Fermi energy can be well labeled by the momentum (correlated band insulator). As seen in Fig 4(a) and (b), in the region between ($k_x, k_y$) =($3\pi/4,0$) and ($\pi/2, \pi/2$), each orbital contributes to the coherence peaks, which is consistent with the behavior of the LDOS (see Fig. 3). Thus, we can conclude that the gap is generated by the SO interaction in this region, while as seen in Fig. \ref{fig:Ak}(c) and (d), in the trivial AF phase, the peaks below the Fermi energy are mainly contributed by the state $(\alpha, \sigma)=(2,\uparrow)$, and thus, the gap is induced by the AF order. This confirms that the topological AF phase is induced by the electron correlation.

\begin{figure}[!h]
\begin{center}
\includegraphics[width=60mm]{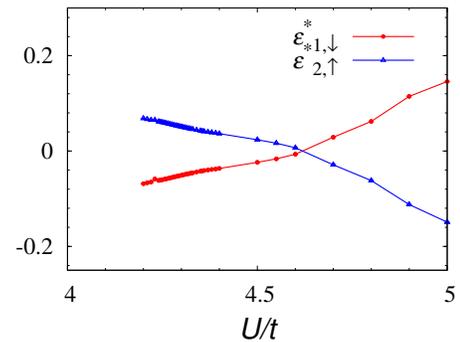}
\end{center}
\caption{
The renormalized energy level of each orbital ($\varepsilon_{\alpha,\sigma}^{*}$) as a function of interaction.
}
\label{fig:renormalized}
\end{figure}

Since the system behaves as a correlated band insulator in the non-trivial AF region, as mentioned above, we expand the self-energies around $\omega=0$ and plot the renormalized energy level of each orbital defined by $\varepsilon_{\alpha,\sigma}^{*}=z_{\alpha,\sigma}(\varepsilon_{\alpha,\sigma}+\Sigma_{\alpha,\sigma}(\omega=0))$, where the $z_{\alpha,\sigma}$ is the renormalization factor of each orbital in Fig. \ref{fig:renormalized}. Note that as long as the system is recognized as a correlated band insulator, the relation $\varepsilon_{2,\uparrow}^{*}>\varepsilon_{1,\downarrow}^{*}$ is the condition required for existence of the SO induced gap. In this figure, we can see that for $4.2<U<4.6$, the state $(\alpha,\sigma)=(1,\downarrow)$ is located below the Fermi energy, implying that the energy gap is dominated by the SO interaction. Since such an SO induced energy gap in the AF phase is not observed at the Hartree-Fock level, we conclude that correlation effects renormalize each band and thus suppress the energy shifts resulting from the spin polarization. This kind of renormalization should commonly occur in TBIs.
Before summarizing this paper, the effects of spatial fluctuations, which are neglected in the DMFT framework, should be mentioned. Spatial fluctuations generally suppress the magnetic order and have a tendency to shift the transition point to the strongly correlated region. Therefore, such fluctuations would usually cause substantial effects on magnetic transitions, so we have to carefully examine our results on the non-trivial magnetic phase. Although we cannot draw a definite conclusion beyond the DMFT results in this paper, we expect that the non-trivial phase found here can persist even when such spatial fluctuations are incorporated. First, we should note that while the magnetic transition point is shifted to the correlated region, the topological properties also remain up to, at least, antiferromagnetic transition point as long as singularities in the Green's function are absent. Furthermore, as demonstrated in this paper, the band renormalization effects, which mainly originate from the local fluctuations, are essential for stabilizing the non-trivial magnetic phase. We thus expect that the essential properties of the non-trivial magnetic phase can be captured with the DMFT treatment, and the qualitative properties may not be changed even in the presence of spatial fluctuations. To confirm this point, however, the microscopic analysis taking into account spatial fluctuations properly (e.g. with cluster extension of DMFT, variational cluster approach, etc.) should be necessary, which is left for our future work.

Finally, we comment on the edge states. The nonzero spin Hall conductivity in the AFTI phase implies the existence of gapless spin excitations on open edges, which carries the spin Hall current, when the system has open boundaries\cite{Sheng05}.  Thus one can deduce that the AF order is suppressed at the edges, and the helical edge states are topologically protected against magnetic instability, in spite of the existence of the bulk AF order.

\section{Summary}\label{sec. summary}

In summary, we have studied topological properties of the AF phases in the extended BHZ model including local Coulomb interaction. The DMFT+NRG calculation of the magnetic moment and the spin Chern number has suggested the existence of a topologically non-trivial antiferromagnetic phase, where one can observe both the quantized spin Hall conductivity and the magnetic order, even if the Hatree-Fock treatment does not support it. We have demonstrated that the correlation effects are essential to realize the non-trivial AF magnetic phase; the correlation effects strongly renormalize the energy-level shift induced by the AF ordering, keeping the system still in the band-inversion regime. Although the detailed situations should depend on the system under consideration, this kind of renormalization effect is inherent in the TBI caused by the band-inversion mechanism, which encourages us to look for topologically non-trivial phases in the strong correlation regime.

\section{Acknowledgments}

This work is partly supported by JSPS through its FIRST Program, KAKENHI (Nos. 21740232, 20104010, 23102714, 23540406), and the Global COE Program ``The Next Generation of Physics, Spun from Universality and Emergence'' from MEXT of Japan.



\begin{thebibliography}{99}
\bibitem{Footnote_hysteresis_HF} In the coexistence region, which is inherent in the first-order transition, the solution depends on the initial condition imposed for solving the self-consistent mean field equation. The stable phase is determined by examining the internal energy of each solution. That is, both the paramagnetic phase and the AF phase can be the self-consistent solutions in the coexistence region.
\bibitem{Footnote_QSH} As noted in ref. 52, periodic boundary condition and the Ward-Takahashi identity are essential for the relation between the conductivity and the spin Chern number. Both of them are remained in antiferromagnetic phase, and therefore we can conclude that the conductivity is still proportional to the spin Chern number.
\bibitem{Footnote_hysteresis_DMFT} In the coexistence region, the solution depends on initial hybridization functions which are successively improved in the DMFT iteration. Therefore, starting from the weakly (strongly) correlated region, we end up with the paramagnetic (antiferromagnetic) solution in this region, giving rise to hysteresis behavior.
\end{thebibliography}
\end{document}